\documentclass[12pt]{article}
\usepackage{amstex}
\newfont{\bit}{cmbxti10 scaled 1728}
\renewcommand{\baselinestretch}{1.24}
\def\d{\partial}
\def\tha{\vartheta(a -\rho) }
\def\itha{\vartheta(\rho-a) }
\def\sqa{\sqrt{a^2-\rho^2} }
\def\da{\delta(\rho- a)  }

\newcommand{\thy}{\vartheta(a-|y|)}
\newcommand{\sqy}{\sqrt{a^2-y^2} }

\newcommand{\ef}{e_\phi }
\newcommand{\er}{e_\rho }
\newcommand{\rhob}{\bar{\rho}}
\newcommand{\ab}{\bar{a} }
\newcommand{\eb}{\bar{e} }
\newcommand{\xb}{\bar{x} }
\newcommand{\thab}{\vartheta(a -\rhob) }
\newcommand{\sqab}{\sqrt{a^2-\rhob^2} }
\newcommand{\dab}{\delta(\rhob- a) }
\newcommand{\dby}{\delta(a-|y|) }

\newcommand{\poch}[2]{\left( #1\right ) _{#2} }
\begin{document}
\renewcommand{\thefootnote}{\fnsymbol{footnote}}
\newpage
\pagestyle{empty}
\begin{center}
{\LARGE {Boosting the Kerr-geometry \\
  into\\
an arbitrary direction\\
}}
\def\er{e_\rho }
\def\er{e_\rho }
\vfill
{\large
Herbert BALASIN
 \footnote[7]{ e-mail: hbalasin @@ email.tuwien.ac.at}
 \footnote[9]{supported by the APART-program of the Austrian
	Academy of Sciences}
}\\
{\em
 Institut f\"ur Theoretische Physik, Technische Universit\"at Wien\\
 Wiedner Hauptstra{\ss}e 8--10, A - 1040 Wien, AUSTRIA
 }\\[.5cm]
{\em and}\\[.5cm]
{\large Herbert NACHBAGAUER
\footnote[8]{e-mail: herby @@ lapphp1.in2p3.fr}
}\\
{\em{Laboratoire de Physique Th\'eorique}}
{\small E}N{\large S}{\Large L}{\large A}P{\small P}
\footnote{URA 14-36 du CNRS, associ\'ee \`a l'E.N.S. de Lyon,
et au L.A.P.P. (IN2P3-CNRS)\\
\hspace*{0.7cm} d'Annecy-le-Vieux}
\\
{\em Chemin de Bellevue, BP 110, F - 74941 Annecy-le-Vieux Cedex,
France}\\[.5cm]
\end{center}
\vfill
\begin{abstract}
We generalize previous work \cite{BaNa3} on the ultrarelativistic limit
of the Kerr-geometry by lifting the restriction on boosting along the
axis of symmetry. \\
\end{abstract}

\vfill

\rightline{{\small E}N{\large S}{\Large L}{\large A}P{\small P}-A-538/95}
\rightline{TUW 95 -- 17 }
\rightline{August 1995}

\newpage
\renewcommand{\baselinestretch}{1.5}
\small\normalsize
\renewcommand{\thefootnote}{\arabic{footnote}}
\setcounter{footnote}{0}
\newpage
\pagebreak
\pagenumbering{arabic}
\pagestyle{plain}

\section*{\Large\bit 1) Introduction}
\par
Although the Kerr-geometry, when considered classically, represents
a vacuum solution of the Einstein-equations, there have been various
attempts \cite{Israel1,Israel2,Burinskii,Lopez}
to find a singular matter distribution as its source.
The natural candidate for locating such an energy-momentum
distribution is the singular region of the spacetime.\par
Using distributional techniques it is indeed possible to find
a tensor-distribution \cite{BaNa2}, which is supported in the
singular region and which from the distributional viewpoint
represents the right-hand side of the Einstein equations.
A central ingredient in this endeavour is the Kerr-Schild
decomposition of the metric which is the main reason for
the applicability of distributional techniques to the non linear
Einstein equations. Moreover, the flat (background) part of the
decomposition provides us with a natural notion of boosts as being
its associated isometries. However, as already noted in the
Schwarzschild-case \cite{AiSe}, boosting the metric itself does
not produce a sensible result. Therefore the strategy is to shift
from the metric to the energy-momentum tensor which has a
well-defined limit \cite{BaNa1}.
The resulting limit is a pp-wave \cite{JEK}.
Solving the Einstein-equations with this inhomogeneity produces
the so-called Aichelburg--Sexl (AS) geometry describing the
gravitational field of a massless point-particle.
The result is independent of the direction of the boost which is
 due to the spherical symmetry of the original geometry. \par

This is, however, no longer the case for Kerr, which is only
axis-symmetric. In a first step \cite{BaNa3} the authors considered
therefore a boost along the preferred direction, namely the axis of
symmetry. The aim of the present work is to lift this restriction
and to investigate the form of the limit by boosting along an
arbitrary direction. Finally we will discuss the extremal case
where the boost direction becomes perpendicular to the axis of symmetry
in some detail.

\section*{\Large\bit 2) The general boost}

The main ingredient of our approach
is the Kerr-Schild decomposition of the Kerr geometry
\begin{equation}\label{KerrSchild}
g_{ab}=\eta_{ab} + f \>k_a k_b,
\end{equation}
where $\eta_{ab}$ denotes the flat background part, with respect
to which boosts
find their natural home. $k^a$ denotes a geodetic null vector field and $f$
a scalar function.
With respect to Kerr-Schild coordinates \cite{HaEl} $\eta_{ab}$ becomes
manifestly flat and $k^a$ and $f$ are given by ($\rho^2 =x^2 + y^2 $)
\begin{eqnarray*}
k^a=(1,k^i), && k^i = \frac{r\rho}{r^2+a^2} e^i_\rho -
                \frac{a\rho}{r^2+a^2} e^i_\phi  \\
f=\frac{2 m r}{\Sigma} , && \Sigma = \frac{r^4 + a^2 z^2}{r^2}
\end{eqnarray*}
where $r$ is subject to $r^4-r^2 (\rho^2 +z^2-a^2) -a^2 z^2 =0.$
The Ricci-tensor for geometries in the Kerr-Schild class takes
the form
\begin{equation}\label{Ricci}
R^a{}_b = \frac{1}{2} (\d^a\d_c(fk^c k_b) + \d_b\d_c(fk^ck^a) -
                \d^2(fk^a k_b) ),
\end{equation}
which  in the Kerr case gives rise to the distributional
energy-momentum tensor \cite{BaNa2}

\begin{align}\label{KerrEM}
T^a{}_b &= \frac{m\delta(z)}{8\pi}\left\{ \frac{2}{a}\left(
\left[ \frac{a^2\tha}{\sqa^3}\right ] -\frac{\tha}{\sqa} -
\da\right )(dt)^a(\d_t)_b \right. \nonumber\\
&+((\d_t)^a(\ef)_b - (\ef)^a (dt)_b)\left(
2\left [\frac{\rho\tha}{\sqa^3}\right ]
-\frac{\pi}{a}\da\right ) \nonumber\\
&+\frac{2}{a}\left( - \left[\frac{\rho^2\tha}{\sqa^3}\right ]
- \frac{\tha}{\sqa}
 + 2\da\right )(\ef)^a(\ef)_b \nonumber\\
&\left. -\frac{2}{a}\frac{\tha}{\sqa}(\er)^a(\er)_b\right \}
\end{align}

The square-bracket terms in (\ref{KerrEM}) represent distributional
extensions of the corresponding non-locally integrable functions
to the whole of test function space.
Their definition may be exemplified by
$$
\left(\left[ \frac{\tha}{\sqa^3} \right] ,\varphi \right) :=
\int\limits_{\rho\leq a} d^2x\frac{1}{\sqa^3 }(\varphi(x)
-\varphi(a e_\rho)),
$$
where $e_\rho$ denotes the radial unit-vector with respect to
polar coordinates. For a more detailed discussion on the origin of
these terms the reader is referred to \cite{BaNa2}.

Interpreting Kerr-Schild coordinates as being asymptotically at rest
we may rewrite $T^a{}_b$ with respect to an arbitrary Lorentz-frame.
The boost-plane is spanned by the timelike vector $P^a= m(\d_t)^a$
and its orthogonal spacelike counterpart $Q^a= m e^a$. Without loss of
 generality we may take $e^a$ to lie in the {\it x-z} plane
\begin{align}\label{newvars}
&m(e_z)^a =  Q^a \cos\alpha -  m\eb^a \sin\alpha
\nonumber\\
& m(e_x)^a = Q^a \sin\alpha +  m\eb^a \cos\alpha
\end{align}
where $\alpha$ denotes the angle between the axis of symmetry
and the direction of the boost and $\eb^a$ the spacelike direction,
which spans together with $(e_y)^a$ the two-plane orthogonal to the
boost. With respect to (\ref{newvars}) the $\delta(z)$ factor in
(\ref{KerrEM}) fixes $(Q x)= m\xb \tan\alpha,\,\xb = ( \eb  x) $,
 which in turn implies that
\begin{equation}\label{radID}
\rho^2 = x^2+y^2 =
       \frac{1}{m^2}(  (Qx) \sin\alpha  + m\xb \cos\alpha )^2 + y^2=
       \frac{\xb^2}{\cos^2\alpha} + y^2 =: \rhob^2.
\end{equation}
So one ends up with the following expression for the
energy-momentum tensor
\begin{align}\label{covEM}
T^a{}_b = &  \frac{\delta(Qx-m\xb \> \tan\alpha )}{8\pi\cos\alpha}
        \left\{ -\frac{2}{a}\left(
        \left[ \frac{a^2\thab}{\sqab^3}\right ] -\frac{\thab}{\sqab} -
        \dab\right )P^a P_b \right. \nonumber\\
&+\left( 2\left [\frac{\rhob\thab}{\sqab^3}\right ]
        -\frac{\pi}{a}\dab\right ) \left \{ P^a \>m  (e_{\phi})_b  +
         m (e_{\phi})^a\>  P_b  \right \}\nonumber\\
&+\frac{2}{a}\left( - \left[\frac{\rhob^2\thab}{\sqab^3}\right ]
        - \frac{\thab}{\sqab} + 2\dab\right )   m (e_{\phi})^a\>
	 m (e_{\phi})_b
\nonumber\\
&\left. -\frac{2}{a}\frac{\thab}{\sqab} m  (e_{\rho})^a\> m (e_{\rho})_b
\right \}
\end{align}
where due to the $\delta$-factor in (\ref{covEM})
\begin{align}
m  (e_{\phi})^a &= \frac{1}{\rho}( x\, m (e_y)^a\! -  y\, m (e_x)^a )
	= \frac{1}{\rhob} \left( -y\sin\alpha\>Q^a\!
	- \frac{m}{\cos\alpha}
	(y\cos^2\alpha\>(\eb)^a\! - \xb\>(e_y)^a  ) \right)  \nonumber\\
m  (e_{\rho})^a &=  \frac{1}{\rho}(x\, m (e_x)^a\! + y \, m (e_y)^a)
	  =\frac{1}{\rhob}\left( \xb\tan\alpha\>Q^a\!  +
	m(\xb\>(\eb)^a\! +y\>(e_y)^a ) \right)
\end{align}
Although this expression may look rather unwieldy in comparison with
(\ref{KerrEM}) it allows a simple ultrarelativistic limit by letting
$m\to 0$ and replacing $P^a$ and $Q^a$ by their null limit $p^a$.
\begin{align}\label{urEM}
T^a{}_b =
& \frac{\delta(px)}{8\pi\cos\alpha}
        \left\{ -\frac{2}{a}\left(
        \left[ \frac{a^2\thab}{\sqab^3}\right ] -\frac{\thab}{\sqab} -
        \dab\right ) \right. \nonumber\\
&+\left( 2\left [\frac{\rhob\thab}{\sqab^3}\right ]
        -\frac{\pi}{a}\dab\right )\frac{2y\sin\alpha}{\rhob} \nonumber\\
&+\left( - \left[\frac{\rhob^2\thab}{\sqab^3}\right ]
         + 2\dab\right )\frac{2y^2\sin^2\alpha}{a\rhob^2}
        \nonumber\\
&\left. -\frac{\thab}{\sqab}\frac{2\sin^2\alpha}{a}  \right \}p^a p_b
	=: -\frac{1}{16\pi}g(\xb,y) \delta(px) \, p^a p_b
\end{align}
As expected the resulting energy-momentum tensor is that of a
pp-(shock)wave. For $\alpha\to 0$ only the first term in the curly bracket
survives and (\ref{urEM}) coincides with the result obtained in
\cite{BaNa3}.

\section*{\Large\bit 3) Perturbative evaluation}
In order to find the metric corresponding to the distributional
energy-momentum tensor (\ref{urEM}), one has to solve the
Einstein equations that  in this setting take the form of the Poisson
equation
\begin{equation}\label{Poisson}
( \d_{\bar x}^2 + \d_{y}^2 ) f(\bar x,y) = g(\bar x,y)
\end{equation}
for the profile-function $f(x)\, \delta(px)$
of the pp-wave. It can be solved straight-forwardly  in a
perturbative way except for the particular case  of the orthogonal
boost $\alpha\to\pi /2$ which needs special care and will therefore
 be dealt with in the next chapter.
Rescaling $\xb$ by $\cos\alpha$ and denoting the new variable by $x$
(\ref{Poisson}) becomes
\begin{equation}\label{genPERT}
(\Delta + \tan^2 \!\! \alpha \; \d_x^2\; )\sum\limits_{n=0}^\infty
	f_n \sin^n\alpha =\frac{1}{\cos\alpha}(g_0 + \sin \alpha\: g_1 +
         \sin^2\! \alpha\: g_2)
\end{equation}
where the $g_i$ may be read off from (\ref{urEM}).
Expanding $\cos\alpha$ and $\tan\alpha$ into power series with respect
to $\sin\alpha$ and grouping corresponding powers together yields
\begin{align}
&\Delta f_0 = g_0,\qquad \Delta f_1 = g_1,\nonumber\\
&\Delta f_{2n} + \sum\limits_{k=0}^{n-1} \d_x^2 f_{2k} =
	\frac{\poch{1/2}{n}}{n!} g_0+
	\frac{\poch{1/2}{n-1}}{(n-1)!} g_2, \qquad n\geq 1,\nonumber\\
&\Delta f_{2n+1} + \sum\limits_{k=0}^{n-1} \d_x^2 f_{2k+1}=
	\frac{\poch{1/2}{n}}{n!}
	 g_1,\qquad n\geq 1.
\end{align}
Let us explicitly derive the first  order perturbation $f_1$
which is determined by
\begin{align}\label{pert}
&\Delta f_1(x) = -8\frac{y}{\rho}\left(
	\left[\frac{\rho\tha}{\sqa^3}\right ]
	-\frac{\pi}{2a}\da\right ).
\end{align}
In the region $0< \rho < a$ the classical analogue of (\ref{pert})
may be separated employing polar coordinates. Decomposing $f_1(x)$
into $\tilde{f}_1(\rho)\sin\phi$ we obtain the radial equation
\begin{align}
&\frac{1}{\rho}\d_\rho(\rho\d_\rho\tilde{f}_1 ) -
	\frac{1}{\rho^2}\tilde{f}_1 =
	-\frac{8\rho}{\sqa^3}\nonumber\\
\intertext{which may be simplified by replacing
	$\rho$ by $a e ^u$}\nonumber
&(\d_u^2 -1)\tilde{f}_1 = -\frac{8e^{3u}}{\sqrt{1-e^{2u}}^3}.
\end{align}
This equation is easily solved by using the Green-function
$$\vartheta(u)\sinh u = \frac{1}{2}\vartheta(\rho-a)
	\left( \frac{\rho}{a} -\frac{a}{\rho}\right )
$$
that gives rise to the particular solution
$$
f_1(x)=\frac{8}{\rho}\sqa\sin\phi.
$$
Taking into account the distributional identities
\begin{align}
\Delta\left( \frac{\tha}{\rho}\sqa \sin\phi \right ) &=
	-\left[\frac{\rho\tha}{\sqa^3}\right ]\sin\phi +\nonumber\\
	&\hspace*{2cm}\frac{\pi}{2a}\da\sin\phi +
	2\pi a\d_y\delta^{(2)}(x)\nonumber\\
\Delta\left( \frac{1}{\rho}\sin\phi\right)&=2\pi \d_y \delta^{(2)}(x),
\end{align}
we find
\begin{equation}\label{perSOL}
f_1(x) = \tha\left(\frac{8}{\rho}(\sqa -a) \right )
	\sin\phi - 8\itha\, \frac{a}{\rho}\sin\phi.
\end{equation}
The choice of the solution is dictated by the distributional
nature of the inhomogeneity and natural boundary conditions
for $\rho\to\infty$ which ensure that $f_1$ vanishes asymptotically,
which is equivalent to the fact that the whole solution tends
to the AS-geometry at large distances.
 The last property may also be derived from the fact that $f_1$ tends to
zero in the limit $a\to 0$. Comparing (\ref{perSOL}) with the result of
the boost along the axis of symmetry \cite{BaNa3} shows that the
rotational contributions being inverse powers
die off only in the limit and not at some finite value. This behaviour is
easily understood
by interpreting (\ref{Poisson}) as a 2-dimensional electrostatic problem.
In the spherically symmetric case the corresponding charge distribution
produces only a monopole momentum and may therefore be replaced by
a pointlike distribution of total charge outside its support.

\section*{\Large\bit 4) Transversal Limit}
In order to calculate the limit $\alpha\to \pi/2$ of (\ref{urEM}) one has to
evaluate the expression on an arbitrary test function $\varphi$.
Only two different types of expressions occur in the calculation, namely
those which arise from the square-bracket terms and the concentrated
(delta) contributions respectively.
In the following we will exemplify the respective limits on typical
representatives from each class.
\begin{align}\label{example}
&\lim_{\alpha\to \pi/2}\frac{1}{\cos\alpha}
	\left( \left[ \frac{\thab}{\sqab^3}
	\right ],\varphi \right ) =
	\lim_{\alpha\to \pi/2}
	\left( \left[ \frac{\tha}{\sqa^3}
	\right ],\tilde{\varphi} \right )= \nonumber\\
&\lim_{\alpha\to \pi/2}\int\limits_{\rho\leq a} d^2 x
	\frac{1}{\sqa^3}\left( \varphi( \cos\alpha  x,y) -
	\varphi(  \cos\alpha a\cos\phi,a\sin\phi)\right)=\nonumber\\
&\int\limits_{\rho\leq a} d^2 x\frac{1}{\sqa^3}\left( \varphi(0,y) -
	\varphi(0,a\sin\phi)\right),
\end{align}
where $\tilde{\varphi}(x,y):=\varphi(\cos\alpha x,y)$.
Unfortunately the above result is not in a very useful form.
Further simplification of (\ref{example}) may be achieved by
integrating out
$x$ in the first term, and $\rho$ in the second.
However, in order to perform these integrations we have to restrict the
domain of integration to a disk of radius $\ab <a$ and do the limit
$\ab\to a$ in the end. More explicitely we find
\begin{align*}
&\int\limits_{\rho\leq\ab} d^2 x\frac{1}{\sqa^3}
	\varphi(0,y)=
	\int\limits_{-\ab}^{\ab} dy\varphi(0,y)
	\int\limits_{-\sqrt{\ab^2-y^2}}^{\sqrt{\ab^2-y^2} } dx
	\frac{1}{\sqa^3}=\\
& \frac{2}{\sqrt{a^2-\ab^2}} \int\limits_{-\ab}^{\ab} dy \varphi(0,y)
        \frac{\sqrt{\ab^2-y^2 } }{a^2-y^2}\\
\intertext{and}
&\int\limits_{\rho\leq\ab} d^2 x\frac{1}{\sqa^3}
	\varphi(0,a\sin\phi)=
	\int\limits_0^{\ab} \frac{\rho d\rho}{\sqa^3}
	\int\limits_0^{2\pi} d\phi \varphi(0,a\sin\phi)=\\
&\left(\frac{1}{\sqrt{a^2-\ab^2}}-\frac{1}{a} \right )
	2\int\limits_{-a}^a\frac{dy}{\sqy}\varphi(0,y).
\end{align*}
Using l'Hospital's rule (\ref{example}) becomes
\begin{align*}
&\frac{2}{a} \int\limits_{-a}^{a} \frac{dy}{\sqy}\varphi(0,y)-
	2 \lim_{\ab \to a}
	\sqrt{a^2-{\ab}^2}\int\limits_{-\ab}^{\ab}
	\frac{dy}{\sqrt{\ab^2-y^2}(a^2-y^2)}\varphi(0,y)=\\
&\frac{2}{a}\int\limits_{-a}^{a} \frac{dy}{\sqy}\varphi(0,y)
	- \frac{\pi}{a}(\varphi (0,a) + \varphi (0,-a))=\\
&\left(\frac{2}{a}\frac{\thy}{\sqy}\delta(\xb),\varphi\right ) -
	\frac{\pi}{a}\left(\dby \delta(\xb) ,\varphi\right).
\end{align*}
The limit of the simplest concentrated contribution gives
\begin{align}\label{example1}
&\lim_{\alpha\to \pi /2}\frac{1}{\cos\alpha}\left( \dab,\varphi\right)=
\lim_{\alpha\to \pi /2}	\left(\da,\tilde{\varphi} \right)=\nonumber\\
&\lim_{\alpha\to \pi /2}
a \int\limits_0^{2\pi} d\phi \varphi(\cos\alpha a\cos\phi,a\sin\phi)=
	a\int\limits_0^{2\pi} d\phi \varphi(0,a\sin\phi)=\nonumber\\
&2a\int\limits_{-a}^a \frac{dy}{\sqy}\varphi(0,y) =
	2a\left(\frac{\thy}{\sqy}  \delta(\xb) ,\varphi \right)
\end{align}
Dealing with the remaining terms of (\ref{urEM}) in the same way we
finally end up with the ``transversal'' energy-momentum tensor
\begin{align}\label{orthEM}
T^a{}_b= & \delta(px) \delta(\xb)\left\{ \frac{1}{2}\dby
-\frac{1}{2}\left(\frac{y}{a}\right )\dby\right\}p^ap_b = \nonumber\\
& \delta(px) \delta(\xb)\delta(y+a)p^ap_b .
\end{align}
It is interesting to note that all contributions localized on the
line segment $|y| \le a$ compensate each other and the energy-momentum
tensor turns out to be concentrated on a pointlike region only.
The corresponding profile function is obtained by solving the
Poisson equation with (\ref{orthEM}) as inhomogeneity, which
gives
\begin{equation}\label{transPRO}
f(x)= -8\log\left( \frac{\sqrt{\xb^2 + (y+a)^2} }{\rho_0}\right),
\end{equation}
where $\rho_0$ denotes the length-scale of the AS-geometry \cite{BaNa3}.
The limit $a\to 0$  in fact reproduces the
AS-profile function.
\newpage
\section*{\Large\bit 5) Conclusion}
In the present paper we showed how to calculate the ultrarelativistic
limit of the Kerr-geometry without putting any restriction on the
direction of the boost. Our method is based upon the energy-momentum tensor
of the original Kerr-geometry, which in contrast to
metric admits a well-defined limit, thus avoiding possible
ambiguities arising from the removal of the infinities
of the metric-limit.
The resulting energy-momentum tensor has the form of a pp-(shock)wave,
depending parametrically on the angle $\alpha$
between the boost direction and the axis of rotation.
This dependence turns the support  of the energy-momentum tensor into
an elliptical region in the two-dimensional subspace of the $px=0$-plane.
Therefore only the limiting cases $\alpha=0$ and $\alpha=\pi/2$,
where the support becomes a circle and a line-segment respectively,
admit a solution in closed form.
Nevertheless the general case allows a perturbative treatment
if one suitably rescales the coordinates and expands the resulting
expression with respect to $\sin\alpha$. An explicit calculation
shows that the ''screening'' behaviour of the longitudinal ($\alpha=0$)
case, where the solution turned into that of AS outside the disk with radius
$a$,
gets modified by contributions such that the whole solution displays only
asymptotic AS-behaviour.
The perturbative expansion breaks down in the limiting case $\alpha\to \pi/2$
where the direction of the boost becomes perpendicular to the axis of symmetry.
Taking into account the distributional nature of the limit it is nevertheless
possible to calculate the profile function of the limiting case in closed form.
\par\noindent
It would be interesting to investigate the dependence of particle scattering
on the angle $\alpha$ in comparison to the $\alpha=0$ case, since the latter
displays exactly the AS-behaviour outside the disk with radius $a$.
Work in this direction is currently under progress.

\vfill
\end{document}